\documentclass[sigconf]{acmart}
\usepackage{makecell}
\usepackage{hyperref}
\usepackage{array}
\usepackage{booktabs}
\usepackage{tikz}
\usepackage{xspace}
\usepackage[normalem]{ulem}
\usepackage{tabularx}
\usepackage{cancel}
\usepackage{graphicx}

\newcommand\jw[1]{#1}
\newcommand\dg[1]{#1}

\newcommand\fix[1]{#1}
\newcommand\need[1]{#1}

\renewcommand{\sout}[1]{}

\newcolumntype{Y}{>{\raggedright\arraybackslash}X}

\AtBeginDocument{%
  }

\setcopyright{acmlicensed}
\copyrightyear{2026}
\acmYear{2026}
\setcopyright{cc}
\setcctype{by}
\acmConference[CHI EA '26]{Extended Abstracts of the 2026 CHI Conference on Human Factors in Computing Systems}{April 13--17, 2026}{Barcelona, Spain}
\acmBooktitle{Extended Abstracts of the 2026 CHI Conference on Human Factors in Computing Systems (CHI EA '26), April 13--17, 2026, Barcelona, Spain}
\acmDOI{10.1145/3772363.3798682}
\acmISBN{979-8-4007-2281-3/2026/04}

\begin{document}

\title{One Is Not Enough: How People Use Multiple AI Models in Everyday Life}

\author{Seunghwa Pyo}
\authornote{These authors contributed equally to this research.}
\affiliation{%
  \institution{Department of Industrial Design, KAIST}
  \city{Daejeon}
  \country{Republic of Korea}
}
\email{shpyo@kaist.ac.kr}

\author{Donggun Lee}
\authornotemark[1]
\affiliation{%
  \institution{Department of Industrial Design, KAIST}
  \city{Daejeon}
  \country{Republic of Korea}
}
\email{jlee4330@kaist.ac.kr}

\author{Jungwoo Rhee}
\authornotemark[1]
\orcid{0009-0005-3832-2835}
\affiliation{%
  \institution{Department of Industrial Design, KAIST}
  \city{Daejeon}
  \country{Republic of Korea}
}
\email{jwoorhee@kaist.ac.kr}

\author{Soobin Park}
\affiliation{%
  \institution{Department of Industrial Design, KAIST}
  \city{Daejeon}
  \country{Republic of Korea}
}
\email{soobinpark@kaist.ac.kr}

\author{Youn-kyung Lim}
\affiliation{%
  \institution{Department of Industrial Design, KAIST}
  \city{Daejeon}
  \country{Republic of Korea}
}
\email{younlim@kaist.ac.kr}



\begin{abstract}

People increasingly use multiple Multimodal Large Language Models (MLLMs) concurrently, selecting each based on its perceived strengths. This cross-platform practice creates coordination challenges: adapting prompts to different interfaces, calibrating trust against inconsistent behaviors, and navigating separate conversation histories. Prior HCI research focused on single-agent interactions, leaving multi-MLLM orchestration underexplored. Through a diary study and semi-structured interviews ($N=10$), we examine how individuals organize work across competing AI systems. Our findings reveal that users construct primary and secondary hierarchies among models that shift over usage context. They also develop personalized switching patterns triggered by task aggregation to adjust effort and latency, and output credibility. These insights inform future tool design opportunities, supporting users to coordinate multi-MLLM workflows.
\end{abstract}




\maketitle

\section{Introduction}

\need{AI assistants} have become ubiquitous productivity tools, with ChatGPT\footnote{\url{https://chatgpt.com/}}, Gemini\footnote{\url{https://gemini.google.com/}}, and Claude\footnote{\url{https://claude.ai/}} now supporting diverse tasks including writing~\cite{DangCHI2024}, programming~\cite{SarkarCHI2024}, and health information seeking~\cite{YunJMIR2025}.  
As these Multimodal Large Language Models (MLLM)---AI systems capable of processing text, images, and code---have matured, people have begun adopting multiple platforms concurrently rather than relying on a single provider~\cite{imagining2025close}. 
This multi-MLLM practice arises because each system offers distinct strengths: individuals select specific models for particular capabilities, such as one for logical reasoning and another for creative generation~\cite{ZamfirescuCHI2023, JonasIJHCI2025}. 
\need{Strategically utilizing specific model capabilities has emerged as a practical skill for optimizing individual workflows and achieving high-quality outcomes~\cite{JonasIJHCI2025, DangCHI2024}.
With MLLMs evolving rapidly and multi-platform use becoming routine, users increasingly mix and match models within a single task to leverage their complementary strengths~\cite{hou2025llmapplicationscurrentparadigms, MicrosoftNFW2025}. 
Yet coordinating across multiple competing MLLM platforms introduces cognitive overhead that existing research has not fully addressed~\cite{JokelaCHI2015}.}

\begin{table*} [!t]
\centering
\caption{Self-reported Participant Demographics and MLLM Models Used}
\Description{This table lists the demographic information and AI usage profiles of 10 participants (P1--P10). Participants are aged 23 to 29, consisting of 6 females and 4 males. Occupations include graduate students in fields like Industrial Design and Computer Science, as well as professionals like developers and designers. All participants use multiple MLLMs, with ChatGPT and Gemini being the most common, followed by Claude and Perplexity. Some also utilize specialized tools like Cursor and Adobe Firefly.}
\label{tab:participants}
\footnotesize
\resizebox{\linewidth}{!}{%
\begin{tabular}{c c c l l l}
\toprule
\textbf{ID} & \textbf{Age} & \textbf{Gender} & \textbf{Occupation} & \textbf{MLLM Models Used} & \textbf{Other AI Services Used} \\ \midrule
P1 & 26 & M & Graduate student in Computer Science & ChatGPT, Gemini, Perplexity & Cursor\\
P2 & 27 & M & Wildlife photographer & ChatGPT, Gemini, Claude & \\
P3 & 29 & F & Graduate student in Industrial Design & ChatGPT, Gemini, Claude & \\
P4 & 25 & F & Graduate student in Electrical Engineering & ChatGPT, Gemini & \\
P5 & 23 & F & Graduate student in Electrical Engineering & ChatGPT, Gemini, Perplexity & Copilot\\
P6 & 25 & F & Engineer / Software Developer & ChatGPT, Gemini, Claude, Samsung Gauss & \\
P7 & 27 & F & Visual Development Designer & ChatGPT, Gemini, Claude & Copilot, Adobe Firefly, Freepik \\
P8 & 26 & M & Full-stack Developer & Gemini, Claude & Antigravity \\
P9 & 27 & M & Graduate student in Industrial Design & ChatGPT, Gemini & \\
P10 & 26 & F & Graduate student in Industrial Design & ChatGPT, Gemini, Claude & \\\bottomrule
\end{tabular}%
}
\end{table*}

\dg{This work builds on prior HCI research on information ecologies, multi-device interaction, and tool switching, while focusing on how these dynamics are reconfigured when users coordinate multiple generative AI systems.} Prior work on multi-device ecologies shows that people face friction when transferring context across platforms~\cite{JokelaCHI2015}.
This context-delegation \fix{between platforms} challenge intensifies with AI \need{systems}, which require users to adapt to prompting styles~\cite{ZamfirescuCHI2023}, calibrate expectations against inconsistent behaviors~\cite{LugerCSCW2016}, and manage divergent conversation histories~\cite{LiaoCHI2020, SuhUIST2023}.  
Foundational HCI research has examined trust calibration~\cite{KocielnikCHI2019, UenoCHIEA2022} and collaborative dynamics~\cite{GeroCHI2023} within single-agent interactions.
However, these frameworks assume a stable human-AI dyad rather than scenarios in which individuals switch between competing systems~\cite{LindgrenIJCI2025}.
Recent studies of \need{generative AI} adoption in professional contexts focus on individual tool use rather than cross-platform orchestration~\cite{Macy_2024}.
Consequently, we have limited empirical understanding of how people develop strategies to allocate tasks, reconcile conflicting outputs, and maintain coherent workflows across multiple MLLM platforms. 

This study addresses this gap through a \need{four-day} diary study and semi-structured interviews examining how users organize and coordinate multiple MLLMs in everyday practice. 
We contribute: (1) an empirical characterization of the roles and relational hierarchies users assign to different MLLM systems, and (2) a taxonomy of coordination strategies users employ to manage cross-platform workflows. 
These findings offer design implications for tools that support users navigating an increasingly heterogeneous AI landscape. Two research questions guide our inquiries:

\begin{itemize}
    \item \textbf{RQ1)} What mental models do users construct when distributing tasks across multiple MLLMs?
    \item \textbf{RQ2)} What strategies do users develop to coordinate their interactions across multiple MLLM platforms?\
\end{itemize}

\section{Methodology}

\subsection{Diary Study and Post-study Interview}

To examine how people organize and coordinate multiple MLLMs in everyday practice, we conducted a qualitative design study that combines diary documentation with follow-up interviews, following recent methodological \need{frameworks} for evaluating AI tool adoption in-situ~\cite{DearDiary}.
Over four days, participants logged their MLLM use through a web-based diary interface, submitting an entry each time they used an MLLM.
Each entry recorded the model used, the rationale for selection, the prompt content, satisfaction, and emotional state, providing in-the-moment accounts of model coordination \need{(see Appendix~A)}.
\dg{Entries were designed to be lightweight, consisting of six short selection prompts and three follow-up questions probing reasons for choices, capturing contextual usage while enabling revisit during debriefs.} After the diary period, we conducted follow-up interviews (3 in-person, 7 remote) to further examine participants' model choices and workflow strategies.
Interviews were facilitated by two researchers, with one lead researcher attending all sessions to ensure protocol consistency; questions probed perceived capabilities, role assignments, and workflow organization.
\need{Our study was approved by the Institutional Review Board.}


\subsection{Participants}


We recruited ten adults (6 female, 4 male; aged 23--29, $M=26.1$, $SD=1.60$; see Table~\ref{tab:participants}) who met two criteria: (1) regular use of at least two MLLM services and \need{(2) practical experience in using different models complementarily for personal or professional projects.} This ensured that all participants possessed sufficient familiarity with cross-platform coordination.
\need{Participants were compensated approximately \$35 USD in local currency.}


\subsection{Data Collection and Analysis}

We collected 129 diary entries ($M=12.9$, $SD=3.41$) and conducted 10 post-study interviews (duration: $M=34.0$, $SD=3.4$ minutes). Interviews were audio-recorded, transcribed, and translated into English. Three researchers conducted an inductive thematic analysis following Braun and Clarke~\cite{BraunQRP2006}. We iteratively developed a shared codebook, generated expanded codes, and consolidated them into 11 final codes and four subthemes, which were synthesized into two overarching themes: (1) Individual MLLM Hierarchy Structures and (2) Cross-Platform Coordination Strategies. The coding process was iterative, with all authors holding regular discussions to resolve discrepancies and reach consensus.


\section{Findings}

\begin{figure*}[t]
    \centering
    \includegraphics[width=1\linewidth]{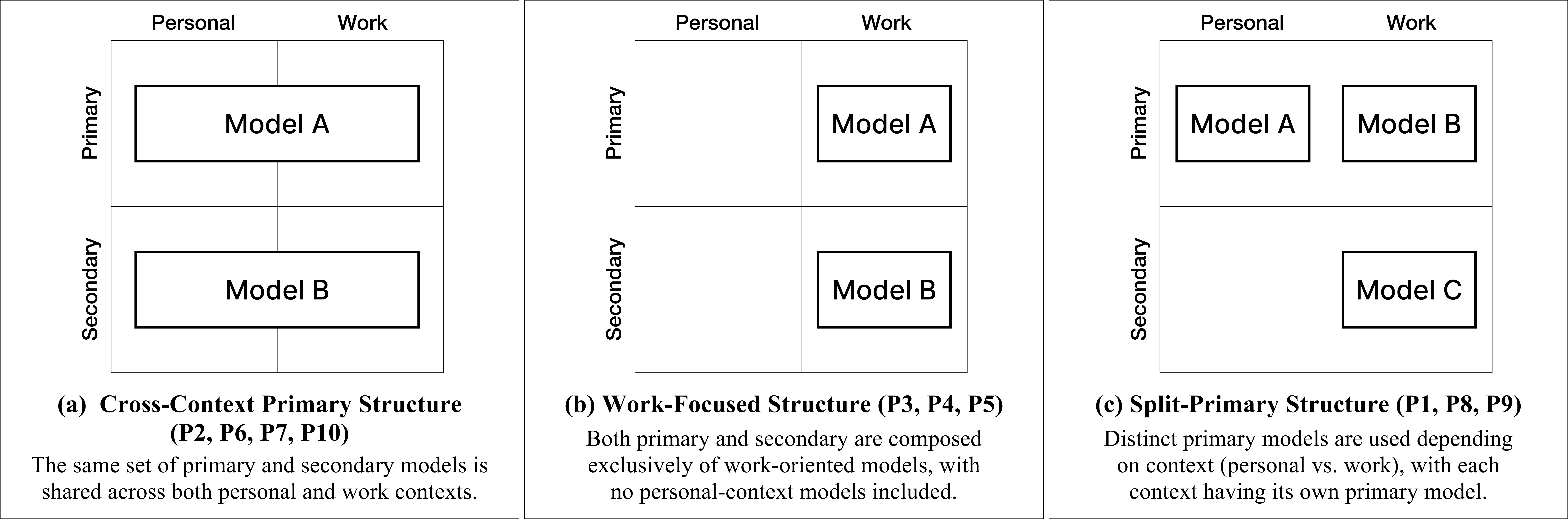}
    \caption{Three recurring hierarchy patterns: (a) Cross-Context Primay, (b) Work-Focused, and (c) Split-Primary.}
    \Description{This figure consists of three 2x2 grids, each illustrating a different structural pattern of how users organize multiple AI models across two dimensions: Importance (Primary vs. Secondary, on the y-axis) and Context (Personal vs. Work, on the x-axis).
    
    (a) Cross-Context Primary Structure: A single box labeled "Model A" spans both Personal and Work columns in the Primary row. Below it, "Model B" spans both columns in the Secondary row. This indicates that the same set of primary and secondary models is shared across both contexts. This pattern is associated with participants P2, P6, P7, and P10.

    (b) Work-Focused Structure: In this grid, the "Personal" column is entirely empty. Both "Model A" (Primary) and "Model B" (Secondary) are located exclusively within the "Work" column. This represents a hierarchy composed only of work-oriented models. This pattern is associated with participants P3, P4, and P5.

    (c) Split-Primary Structure: In the Primary row, "Model A" is in the Personal column, while "Model B" is in the Work column. In the Secondary row, "Model C" is placed under the Work column. This shows that distinct primary models are used depending on the context, with each context having its own primary model. This pattern is associated with participants P1, P8, and P9.}
    \label{fig:figure1}
\end{figure*}

Our findings show that participants employed diverse, personalized ways of using and coordinating MLLMs. In the following section, we present: (1) the hierarchy structures participants formed across models, and (2) the strategies they used to navigate and coordinate multi-MLLM workflows.

\subsection{Individual MLLM Hierarchy Structures}

\subsubsection{Different Hierarchy Across Personal and Work Contexts}
When coordinating multiple MLLMs, \need{all ten} participants chose to utilize the primary model and turned to one or more secondary models as needed. The primary model was used for frequent, core tasks, while secondary models were used for verification \fix{(P2, P3, P5, P7)}, refinement \fix{(P3, P6, P9, P10)}, or more specialized needs \fix{(P2, P4, P5, P6, P7, P9, P10)}. However, its configuration shifted with context—particularly between personal and professional use. Across participants, we identified three recurring hierarchy patterns that describe how model roles were arranged across personal versus work contexts (see Figure~\ref{fig:figure1}).

\textbf{(a) Cross-Context Primary Structure} 
Four participants (P2, P6, P7, P10) centered their workflows around a single favorite model used across both personal and professional contexts. They perceived the favorite model as a general-purpose assistant and kept interactions consolidated. Participants framed the preference around familiarity and accumulated interaction history—describing their favorite model as one that could interpret intent with minimal re-explanation. P6, for example, used ChatGPT for both emotional support and drafting workplace reports:
\textit{``I prioritize it because of the familiarity... when I ask a question, it already knows what I mean. I don't have to explain myself too much.''}
While these participants primarily stayed within their go-to model, they occasionally brought in secondary models when they wanted a second opinion or when a task called for capabilities they associated more strongly with another system.

\textbf{(b) Work-Focused Structure} 
However, three participants (P3, P4, P5) kept one primary model dedicated to professional tasks, treating it as a productivity tool. They deliberately did not feel a strong need to use this primary model in personal contexts. 
P5 \fix{said}: \textit{``I never use it for personal counseling because I don't think it would be helpful... it just feels like a waste of time.''}
Instead, they brought in secondary models mainly to improve work efficiency—for example, to validate outputs, compare responses across models, and catch errors during professional workflows. 
P3 \fix{added}: \textit{``I paste what Claude said into ChatGPT and ask: `Claude thinks this, what do you think?' to provoke a debate. I want them to discuss and find a compromise.''}

\textbf{(c) Split-Primary Structure} 
Furthermore, three participants (P1, P8, P9) maintained distinct primary models for personal versus professional contexts. They typically used a lighter or faster option for low-stakes personal tasks, while reserving a stronger model for work tasks that demanded higher reliability.
P1 \need{explicitly separates his tools based on the cost of failure}: \textit{``For casual queries like pharmacy recommendations or vehicle maintenance, I use Gemini Fast because it's the quickest and doesn't require reasoning. But for work, I strictly use the highest-spec model (GPT-5.2) because fixing errors from a cheaper model wastes more of my time than the token cost.''} 
For these participants, secondary models were primarily considered for task-oriented work needs, whereas personal use tended to remain within the chosen personal primary model.
P8 \fix{elaborated}: \textit{``I switch based on the specific feeling I remember from past usage. If the task requires grasping long contexts, I use Gemini. But if the logic is complex, Claude feels smarter, so I use Claude specifically for that.''}


\subsubsection{Factors Shaping Model Hierarchies}

Across our study, participants formed model hierarchies, with the particular MLLM occupying each role continually shaped by (1) first impressions, (2) expert consensus and social signals, and (3) costs of model use, as participants re-evaluated their options.

\textbf{First-Impression Lock-in}
Early experiences with a model often set a reference point that continued to shape later preferences (P3, P4, P8). For instance, P3 established Claude as their primary writing agent based on a strong initial impression, contrasting its calm style with ChatGPT's performative tone:
\textit{``ChatGPT felt like it was just showing off with fancy words... whereas Claude felt calm, concise, and clear. Since that moment, that impression became solidified, so I still maintain high trust in Claude for writing.''}
This lock-in was especially pronounced when the model’s style or reasoning process felt aligned with how participants approached the task. P2, a wildlife photographer, favored Gemini because its image analysis process matched their birdwatching expertise:
\textit{``Gemini classifies birds exactly like a birdwatcher would, checking critical points like the beak shape or shin feathers first... The way it structures its observation matches my own thinking process.''}

\textbf{Expert Consensus and Social Signals}
Some participants (P8, P9) stayed attentive to which models were perceived as improving or leading, using expert opinions and community discussions as cues for when to re-evaluate their tool choices. P9 treated expert consensus as a strategic filter, monitoring the market without the burden of constant hands-on testing:
\textit{``I can never know better than those AI researchers... So I just trust their reviews. If they say on LinkedIn that a model is rising, that becomes my only standard.''} 
Others (P2, P3, P7, P10) reported trying models after hearing positive recommendations from people around them. 
In contrast, P1 deliberately bypassed social trends and relied strictly on technical performance evidence:
\textit{``I don't look at social reactions. I only trust objective benchmarks like SWE-bench... If a new model scores higher, I switch immediately because relying on an inferior tool wastes my time.''}

\textbf{Costs of Model Use}
Model preferences were sometimes shaped less by perceived performance, as they balanced task suitability against recurring subscription fees or usage costs (P1, P5).
For example, P1 offloads trivial daily queries to Gemini Fast Mode to minimize operational costs,
noting that the financial burden of his high-frequency usage necessitates strict budget management
\textit{``My usage volume is huge... I already spend over 10 dollars a day. If I calculate that monthly, it is an enormous amount, so cost is a very critical factor in my choices.''}

\subsection{MLLM Coordination Strategies}

\subsubsection{Switch Model with Purpose}


Participants developed strategies to coordinate their interactions across multiple MLLM platforms, switching intentionally by assigning different task stages to different models, adjusting effort and latency to task demands, and cross-checking credibility-critical information.

\textbf{Assigning Models to Specialized Roles}
A common coordination strategy was to break a task into stages and assign each stage to the model perceived as strongest for that subtask (P2, P5, P6, P7, P9)—for example, ideation with one model, drafting with another, and polishing or refinement with a third.
P6 described this transition for their SOP writing workflow:
\textit{``I initially brainstormed all my ideas with ChatGPT... it had really nice suggestions. But for the writing style, I took the same essay from ChatGPT and put it in Claude for the polishing.''}
This sequential switching reflected a deliberate choice to leverage different models’ strengths at different points in the workflow.

\textbf{Managing Effort by Task Difficulty}
Three participants (P1, P5, P9) also switched models based on the scale and difficulty of the immediate task to maximize efficiency.
This was a deliberate choice to avoid over-investing effort or waiting for a slow, high-performance model to complete a trivial job.
P9 switched between modes within a single service to manage latency, treating them as agents with different speeds and ranks:
\textit{``If I require logical flow, I use Gemini Thinking Mode. But to change just one word in the result, I switch to Fast Mode immediately... It feels like handing a trivial task to a faster, lower-ranked entity who doesn't need to think deeply.''}


\textbf{Cross-Checking for Credibility}
When information needed to be reliable, participants (P2, P3, P5, P7) adopted a cross-checking strategy by switching across platforms to verify outputs and reduce hallucination risk. 
This strategy was often implemented as a simple one-step verification: participants took the initial output to another model to confirm key claims. 
While many participants relied on this single-step check, P7 used a three-stage sequence (ChatGPT $\rightarrow$ Gemini $\rightarrow$ ChatGPT) to locate evidence and then return to the original model to assess validity:
\textit{``I paste ChatGPT's output into Gemini and ask it to find supporting evidence. Then, I feed that evidence back into ChatGPT and ask: `Is this evidence actually correct?'... It is essentially a three-stage validation process.''}

\subsubsection{Don't Migrate, Iterate Instead}
Rather than migrating across platforms, some participants sought to improve output quality by iterating with a single model. They described two reasons for this choice: (1) the model already knew their context, making switching costly; and (2) when outputs fell short, they sometimes located the source of failure in their own input specification rather than model capability, leading them to revise and clarify their requests instead of migrating.

\textbf{No Resetting the Relationship}
Four participants (P2, P4, P6, P10) sometimes chose to stick with a familiar model because re-establishing background and personal context elsewhere felt more costly than any likely performance gain.
P2 avoided switching away from Gemini for personal hobbies, noting that retraining a new model on his specific persona would require excessive effort:
\textit{``Gemini knows I'm crazy about birds... but Claude would just think I'm a weirdo. I could theoretically spend a whole day taming Claude to understand my context, but that is just too annoying.''}

\textbf{Blame the Prompt, Not the Model}
When results were unsatisfying, participants (P1, P8, P9) sometimes focused on improving how they prompted rather than blaming the model, iterating until the output met their needs.
P9 described this mindset explicitly, comparing poor prompting to user error in driving:
\textit{``Being unsatisfied after lazy prompting is like a bad driver claiming a Tesla is a bad car... I just need to steer it better.''}
P1 similarly framed prompt refinement as a debugging exercise, iterating on prompts and test cases with the same model until the output met their requirements:
\textit{``If the result is unsatisfactory, it is usually because I gave insufficient information. So I don't switch; I just provide more test cases and grill the model until it passes the criteria.''}

\section{Discussion}
\jw{Multi-MLLM use represents a broader interaction paradigm from single-system workflows~\cite{MicrosoftNFW2025}, demanding coordination practices that prior human-AI interaction research has not fully addressed.
Prior work has primarily investigated how users calibrate trust in, adapt to, or seek explainability from a single AI system~\cite{AmershiCHI2019, LiaoCHI2020, KocielnikCHI2019}.
Our findings reveal that }users treated AI models as an evolving \jw{ecosystem: they construct and maintain priority hierarchies across systems, negotiate context boundaries between models, and route tasks based on dynamic signals rather than fixed preferences.}


\jw{Our findings are also consistent with prior HCI research on multi-device use, particularly studies of device ecologies and information foraging. \citet{JokelaCHI2015} identified four device use patterns—sequential use, resource lending, related parallel use, and unrelated parallel use—grounded in physical device availability and task handoff.
At the same time, our findings suggest important differences in how switching operates across MLLMs. Rather than being driven primarily by physical access, switching in our study often required users to re-establish conversational context, reformulate prompts, and selectively transfer prior exchanges across models. Model roles and hierarchies also remained fluid, as participants re-evaluated them in response to peer recommendations, benchmark narratives, interface changes, and subscription constraints.
While utility-based frameworks could predict that users simply select heavier tools for complex tasks and lighter tools for trivial ones, our data show that this selection logic is neither static nor purely performance-driven.
Hierarchy positions shaped in response to first-impression lock-in, expert consensus cues from professional networks and benchmarks, and subscription cost constraints—factors that utility-maximization models do not capture.
Furthermore, cross-checking behavior is distributed in a multi-MLLM context, unlike in single AI systems~\cite{AmershiCHI2019, UenoCHIEA2022}: users exploit divergent model perspectives to detect hallucinations, a strategy that presupposes a mental model of inter-model disagreement as a reliability signal.
This extends trust-calibration research from a dyadic user-system relationship to a networked, multi-agent verification practice.}

Building on this, we highlight two design opportunities. 
\jw{First, a workflow-aware interface could allow users to transfer a structured task representation between models without losing continuity. Current MLLM interfaces treat each conversation as an isolated session, forcing users to reconstruct task framing when switching between models.
Participants who switched models mid-task—for example, moving from ChatGPT for ideation to Claude for polishing—lost established context and had to re-specify goals and constraints from scratch, creating friction that disproportionately penalizes complex, multi-stage workflows.
A portable \textit{task card} could capture the current task state—including goals, active constraints, and prior model outputs—which the user presents to a new model to resume work without re-grounding.
This feature extends transparency guidelines~\cite{AmershiCHI2019}, which address single-system explainability, toward cross-system continuity as a first-class design concern.}

\jw{Second, model-level memory partitioning to segregate interaction histories by domain or task type within a single interface} could help users to protect distinct model contexts. \jw{Users currently face a binary choice between full history retention and complete erasure, which is poorly suited to the nuanced context-management strategies we observed.
Participants actively worked around this limitation: P10 used ChatGPT's temporary chat feature to prevent casual queries from contaminating primary work history, and P1 routed trivial queries to a secondary model to preserve the context purity of primary coding environment.
A concrete design response would be topic containers (e.g., ``Coding,'' ``Health,'' ``Project 1'') that isolate histories within or across models, enabling selective context portability.
This design would address the mismatches between user mental models and system memory behavior~\cite{LugerCSCW2016}, extending their single-agent framing to multi-model ecosystems where context leakage across models is an additional failure mode.}

\section{Limitations and Future Work}

We acknowledge two primary limitations that motivate future work. 
\jw{First, our ten participants were predominantly in their twenties and technically proficient, which likely inflates the prevalence of benchmark-driven switching strategies and deliberate prompt iteration.
Less expert users may rely more heavily on social signals or default to a single model, and older populations may exhibit different patterns of hierarchy formation; future work should sample across expertise levels and age groups to assess how coordination strategies vary.
Second, the four-day diary window captured routine everyday practices but not the longer-term dynamics most consequential for hierarchy stability—specifically, how users respond to major model releases, pricing changes, or capability regressions.
Longitudinal studies tracking model preferences across several months would reveal whether the hierarchy patterns we identified are durable or represent transient states in a rapidly evolving landscape.}

\section{Conclusion} 
We presented a qualitative study of how users navigate and coordinate across multiple MLLMs. Through a diary study and follow-up interviews, we characterized (1) how participants structured model hierarchies across personal and professional contexts and (2) the coordination strategies they used across platforms. This work is timely because, as MLLM ecosystems evolve rapidly, users increasingly need practical skills for selecting, combining, and validating model outputs in everyday practice. Our findings offer an empirical account of how users orchestrate multiple models, complementing prior human--AI interaction work that has primarily examined single-system use. These results point to design opportunities for helping users adapt to evolving AI ecosystems by supporting effective model selection, cross-model coordination, and reliability-oriented use.

\begin{acks}
This work was supported by the National Research Foundation of Korea (NRF) grant funded by the Korea government (MSIT) (No. RS-2021-NR059056).
\end{acks}

\bibliographystyle{ACM-Reference-Format}
\bibliography{references}


\appendix

\section{Appendix: Diary Study Questionnaire}
\label{appendix:diary-form}

\begin{description}
    \item[Participant ID] [Open-ended text]
    
    \item[MLLM Model Used] 
    Specific model name (e.g., Gemini 1.5 Pro, GPT-4o, Claude 3.5 Sonnet).
    
    \item[Reason for Model Selection] 
    The criteria for choosing the specific model (e.g., accuracy, speed, cost, specific feature support, or comparison with other available models).
    
    \item[Purpose of Use] 
    The primary goal of the interaction, such as:
    \begin{itemize}
        \item \textbf{Work:} Planning, report writing, code review, etc.
        \item \textbf{Consultation/Response:} Customer support scripts, HR, coaching, etc.
        \item \textbf{Study/Research:} Paper summarization, idea validation, etc.
    \end{itemize}

    \item[Input Prompt (Optional)] 
    The representative prompt or a brief summary of the query used.

    \item[Satisfaction (0--10)] 
    A 11-point Likert scale measuring the user's satisfaction with the model's response.

    \item[Reason for Satisfaction/Dissatisfaction] 
    Brief qualitative feedback on the strengths or weaknesses of the interaction.

    \item[Current Emotion] 
    Selection of the user's emotional state following the interaction: 
    \textit{Happy, Calm, Sad, Angry,} or \textit{Other}.

    \item[Reason for Emotion] 
    Contextual reason for the selected emotional state.
\end{description}

\end{document}